# Claiming to Detect AI-Written Texts: An Already Obsolete Conception of Working with LLMs? — Extended Bilingual Research Edition

**Olivier Las Vergnas**

Emeritus Professor, Université Paris Nanterre; associated with Université de Lille

Email: olivier.lasvergnas@parisnanterre.fr

## Abstract

When post hoc AI-text detectors are used to assess intellectual work, the issue is not merely the reliability of their classifications but the relevance of the conception of text production underlying their use. This article argues that this conception, centred on textual provenance, is already obsolete as a general framework for understanding human–LLM collaboration. Drawing on a previously developed framework distinguishing three regimes of generative AI use (prosthesis, orthesis and auxesis), it examines the particular case of deep orthesis, a methodological approach to sustained intellectual cooperation under human control. Different regimes of use can lead to identical final wording: neither finer segmentation nor additional provenance categories can therefore uniquely identify the regime of use from the final text alone. Exploratory trials with Pangram illustrate the context sensitivity of textual attribution but do not constitute a performance benchmark. The article further hypothesises that evaluative uses of such detectors may perpetuate a restrictive representation of what human–machine collaboration can entail. It proposes reflexive documentation of intellectual processes as complementary evidence and outlines a research agenda on the evolution of representations of what can actually be checked, what stakeholders believe can be checked, and what they consider legitimate or necessary to check.

## Publication and edition note

The scientific introduction (including references) precedes the English translation (five pages) and the original French text (five pages). The two language versions retain their pagination and content; their layout has been adapted for this edition. The introduction is not part of the journal publication and introduces no new experimental data. Generative AI assistance is disclosed in the article (Box 2) and in the introduction.

# Research context, epistemological foundations and original contribution

This introductory section clarifies the research context, methodological approach and central argument of the article originally published in *L'Esprit d'Archimède*, no. 16, pp. 31–35 (2026). It reports no new experiments and does not present the original work as a benchmark study.

## 1. Epistemological purpose and status of the observations

This article examines, from a conceptual and epistemological perspective, the appropriateness of using AI-text detection tools to assess the intellectual work involved in text production. It applies a previously developed framework distinguishing different regimes of generative AI use. It does not aim to evaluate detectors systematically or to validate that framework. The discussion sets out the central epistemological critique, clarifies the role of the empirical observations and considers the potential societal implications of treating textual attribution as an indicator of intellectual work.

**The central critique: the object and units of analysis.** Post hoc AI-text detectors infer aspects of textual provenance from the final text, either at the document level or at the level of textual segments. However, cooperation with generative AI encompasses various forms of co-writing and, more broadly, intellectual co-production, including the mode referred to below as *deep orthesis*. In this latter mode, intellectual contributions emerge through sustained exchanges, including the formulation of questions, the construction of arguments and critical validation. Such intellectual activity cannot be adequately characterised solely by assigning provenance categories to segments of the final text.

**The role of empirical counterexamples.** The limited number of Pangram trials reported here serves to challenge specific interpretations of detector outputs, in a spirit of critical testing inspired by Popper (1935/1959, §§ 22 and 66). These trials document cases in which unchanged sentences belong to textual segments assigned different provenance categories across versions of a text. Such observations challenge the interpretation of these classifications as stable indicators of the provenance of individual sentences. They neither provide estimates of detection error rates nor constitute a general falsification of probabilistic detection claims.

**The societal argument.** Treating textual provenance as an indicator of intellectual work risks reinforcing a conception of human–AI relationships centred on who produced the text rather than on how the intellectual work was carried out. Whether and how such evaluative practices shape representations of what can and should be controlled in human–AI cooperation is a research question arising from this argument, rather than a conclusion established by the trials.

## 2. Research context: from textual attribution to intellectual cooperation

A common formulation of AI-text detection is the task of distinguishing human-written from AI-generated text. Evaluations of existing detectors have identified limitations in accuracy and robustness (Weber-Wulff et al., 2023). Research on hybrid writing has extended the scope of detection tasks. Zeng et al. (2024) investigated sentence-level detection in texts produced through human–AI collaborative writing. Zhang et al. (2024) introduced MixSet to investigate the detection of texts involving human or AI revision. Su et al. (2025) investigated word- and sentence-level attribution, while Thai et al. (2026) developed a method for estimating the extent of AI editing.

Recent studies also reconsider how the detection task itself is defined. Cheng et al. (2025) propose distinguishing the roles played by LLMs in text production and estimating their degree of involvement. Dycke et al. (2026) distinguish different definitions of AI-generated text and use detailed production histories to evaluate detectors. Bsharat et al. (2026) investigate how detectability varies with the type and sequence of editing operations. Zhao et al. (2026) propose a classification of AI involvement in writing that distinguishes surface-level revision from interventions affecting discourse structure.

Amin et al. (2026) question the validity of likelihood-based measures of human contribution, demonstrating their sensitivity to the context available to the evaluator and identifying limitations in what can be inferred from recorded interactions alone. Their work reinforces the need to distinguish textual attribution from the assessment of intellectual contribution. Building on these concerns, the present paper examines whether the object and categories of AI-text detection are appropriate for characterising the different regimes of generative AI use in intellectual work.

Geng and Poibeau (2025, Sections 2, 6 and 8) have already questioned the definition of LLM-generated text and warned against reducing questions of authorship and intellectual responsibility to simplified detection categories. The present contribution shares this critical perspective and draws on an existing framework distinguishing three regimes of generative AI use—prosthesis, orthesis and auxesis—jointly developed by Jeunesse and Las Vergnas (Jeunesse, 2025, pp. 150–151) and applied to adult learning (Las Vergnas et al., 2025).

Drawing on this framework of generative AI use regimes, the specific contribution of this paper is to argue that the conception of text production underlying provenance-based post hoc detectors is already obsolete in view of the diversity of possible forms of human–AI collaboration. In particular, this conception does not adequately account for forms of intellectual co-production such as deep orthesis. The use of these detectors to assess intellectual work therefore risks perpetuating a restrictive representation of what human–AI collaboration in text production can entail.

## 3. Scope and interpretation of the exploratory observations

The exploratory Pangram trials are described in the original article (pp. 32–35). Conducted using a black-box commercial service, they differ in the degree to which textual transformations were controlled. The reported percentages are detector outputs rather than independently established proportions of AI-generated text. These observations do not constitute a systematic performance evaluation. The two reports reproduced in the original article identify Pangram 4.1. Its model card identifies AI translations and reference sections as out-of-scope material and warns of possible PDF extraction artefacts (Glickenhaus et al., 2026).

Evidence of high classification accuracy, such as that reported by Jabarian and Imas (2025) for Pangram on their evaluation corpus, is compatible with the present argument. The exploratory observations reported here question certain interpretations of detector outputs, particularly their interpretation as context-independent indicators of textual provenance. They motivate the conceptual inquiry but do not constitute its necessary empirical foundation.

## 4. Construct validity and conceptual obsolescence

Classification accuracy relative to a specified labelling scheme, stability under transformations that preserve the attribute under investigation, and the validity of interpreting detector outputs as indicators of intellectual work are distinct issues. In the tradition of construct validation (Cronbach & Meehl, 1955), and its application to computational systems (Jacobs & Wallach, 2021), such an interpretation requires evidence that the operational indicator adequately represents the construct of interest. Using textual provenance classifications to characterise intellectual work therefore involves an additional inference that requires justification.

Consider two hypothetical researchers. One delegates the construction of an argument to an LLM and accepts it without critical examination; the other develops and critically examines the same argument through sustained dialogue with the LLM. Both subsequently ask the model to formulate the argument and obtain identical final wording. The same text can thus result from different regimes of generative AI use. Neither finer segmentation nor additional provenance categories can resolve this ambiguity on the basis of the final text alone. Moreover, intellectual contributions arising from a dialogue may extend across the entire text without being localisable within individual textual segments.

The diversity of human–AI writing practices is not merely hypothetical. Observations of researchers working with AI writing assistants document uses ranging from idea generation to critical examination and revision (Monge Roffarello et al., 2025). Experimental research on argumentative co-writing also identifies different patterns of appropriation of LLM-generated material, associated with differences in perceived authorship and contribution (Luther et al., 2026). These studies document a diversity of practices, although they do not establish the prevalence of deep orthesis as defined here or its evolution over time.

The notion of obsolescence used here concerns the inadequacy of a conception of intellectual work centred on textual provenance when it is used as a general framework for assessing work involving generative AI. It does not imply that AI-text detection is technically obsolete for every purpose, nor does it presuppose that deep orthesis has become a dominant practice. Investigating the diffusion of such forms of cooperation over time, both in absolute terms and relative to other regimes of use, remains a direction for further empirical research.

## 5. Origins and status of the three-regime framework and deep orthesis

The metaphorical distinction between prosthesis and orthesis predates generative AI. Gobert (2001) discussed it in relation to multimedia technologies, Meller (2019) in relation to AI applications, and Lardellier and Carré (2025) in the context of higher education. These uses of the metaphor differ in their conceptual scope and should not be equated with the three-regime framework used here.

The three-regime framework was jointly developed by Christophe Jeunesse and Olivier Las Vergnas in 2024, during the preparation of Jeunesse's habilitation thesis, which explicitly documents their proposal to introduce auxesis (Jeunesse, 2025, pp. 150–151).

The framework provides an analytical distinction between three regimes of generative AI use. Prosthesis refers to substitution for human activity with limited critical control by the user; orthesis to support for an activity whose objectives, orientation and validation remain under human control; and auxesis to the amplification of productive capacities beyond what users could achieve alone, potentially exceeding their ability to assess the resulting outputs. These regimes characterise ways of using AI rather than intrinsic properties of AI systems.

Building on this framework, Las Vergnas progressively formalised a methodological approach, *deep orthesis*, from early 2026 onwards, presenting it at the third Econom'IA Workshop in Cergy in May (Las Vergnas & Jeunesse, 2026). Whereas the three-regime framework is primarily descriptive and analytical, deep orthesis has a methodological and prescriptive purpose. It aims to maintain human control over intellectual work by structuring and documenting interactions, distinguishing the functions of the tools involved, and organising the verification of intermediate and final outputs.

These regimes are not necessarily mutually exclusive: partial delegation and substantial amplification may coexist with active human control (cf. Shneiderman, 2020). Deep orthesis is a methodological development of the orthesis regime, not a fourth regime. Sustained dialogue does not in itself demonstrate effective human control. The present article mobilises these conceptual and methodological frameworks without claiming to introduce or validate them.

## 6. Institutional reactivity and a prospective research agenda

The intended function of a detector and the function assigned to it in practice must be distinguished. Turnitin (n.d.), for example, explicitly cautions against using its AI-writing report as the sole basis for adverse decisions about students. The criticism developed here concerns evaluative uses in which textual attribution is treated as a privileged proxy for intellectual work, rather than the stated purposes of all detection tools.

Espeland and Sauder (2007) show how public measures can reshape expectations and conduct through processes of reactivity. Sarkar (2025) examines negative judgements associated with AI use as a form of boundary work, including their institutional expression. These perspectives suggest that evaluative uses of AI-text detectors may contribute to shaping the perceived legitimacy of different forms of human–AI cooperation. This is a theoretically informed hypothesis, not an effect established by the exploratory observations.

A research perspective arising from this argument is to study how representations of what can and should be controlled in human–LLM intellectual work evolve. Three dimensions must be distinguished: what can actually be checked, what stakeholders believe can be checked, and what they consider legitimate or necessary to check. Their evolution could be investigated through longitudinal analyses of institutional guidelines and assessment criteria, combined with interviews and observations of evaluators and users.

## 7. From textual attribution to reflexive process documentation

Where the objective is to characterise intellectual work rather than textual provenance, documentation should focus on how the activity is organised. Relevant dimensions include the formulation of questions and objectives, the construction and revision of arguments, the exercise of critical and methodological control, and the validation of knowledge claims. These are illustrative dimensions of intellectual activity, not a new assessment scale or an index of authorship.

Interaction logs, successive versions, source records and accounts of decision-making may help document reflective cooperation. Such records are necessarily partial and do not in themselves establish effective human control or intellectual quality. Reflexive documentation should not be equated with compulsory surveillance: selective documentation may support transparency and critical reflection while respecting confidentiality. The objective is to make the modalities of cooperation intelligible, not to replace one simplistic score with another.

## 8. Limitations, data availability and use of generative AI

The exploratory observations concern selected texts and one commercial detector. Exact inputs, settings, timestamps and repeated runs are not all documented in the original article. These observations do not support generalisations about detection performance across detectors, versions, domains or languages. Neither the institutional effects of detection nor the prevalence of different regimes of generative AI use has been measured in this study.

Box 2 of the original article describes its production in deep-orthesis mode and provides a link to the main conversational thread archived on Nakala: https://nakala.fr/10.34847/nkl.afa74m7q. The version-specific link is provided in the original article. This archive is not intended as a complete replication package.

ChatGPT was used for theoretical and technical discussion, documentary assistance, structuring, linguistic revision, illustration, English translation and preparation of this supplementary introduction. The human author retains full responsibility for the research questions, observations, interpretations, verification of references and final scientific content. AI systems are not listed as authors.

## References for the scientific introduction

The references below complement the bibliography of the original article. Preprints are identified as such; they are not presented as peer-reviewed publications.

Amin, H., Lu, Z., & Yin, M. (2026). *Likelihood is not authorship: Can we reliably measure human contribution in AI-assisted writing?* [Preprint]. arXiv. https://doi.org/10.48550/arXiv.2610.19767

Bsharat, S. M., Liu, J., Zhao, X., Yao, T., Shang, X., Tang, Y., Cui, J., Elhagry, A., Al Khatib, S. K., Li, H., Khan, S., & Shen, Z. (2026). *Operation-guided progressive human-to-AI text transformation benchmark for multi-granularity AI-text detection* [Preprint]. arXiv. https://doi.org/10.48550/arXiv.2606.06481

Cheng, Z., Zhou, L., Jiang, F., Wang, B., & Li, H. (2025). Beyond binary: Towards fine-grained LLM-generated text detection via role recognition and involvement measurement. In *Proceedings of the ACM on Web Conference 2025* (pp. 2677–2688). ACM. https://doi.org/10.1145/3696410.3714770

Cronbach, L. J., & Meehl, P. E. (1955). Construct validity in psychological tests. *Psychological Bulletin, 52*(4), 281–302. https://doi.org/10.1037/h0040957

Dycke, N., Sakharova, M., Daheim, N., & Gurevych, I. (2026). *"Your AI text is not mine": Redefining and evaluating AI-generated text detection under realistic assumptions* [Preprint, v2]. arXiv. https://doi.org/10.48550/arXiv.2606.04906

Espeland, W. N., & Sauder, M. (2007). Rankings and reactivity: How public measures recreate social worlds. *American Journal of Sociology, 113*(1), 1–40. https://doi.org/10.1086/517897

Geng, M., & Poibeau, T. (2025). *On the detectability of LLM-generated text: What exactly is LLM-generated text?* [Preprint]. arXiv. https://doi.org/10.48550/arXiv.2510.20810

Glickenhaus, B., Thai, K., Spero, M., Emi, B., & Aerts, A. (2026, October 19). *Pangram 4.1 model card*. Pangram Labs. https://www.pangram.com/research/model-card/pangram-4-1

Gobert, T. (2001). L'orthèse multimédia. In J.-M. Berthelot (Ed.), *Recherches en sciences sociales : Jalons et segments* (pp. 135–154). L'Harmattan. https://hal.science/hal-05097774v1

Jabarian, B., & Imas, A. (2025). *Artificial writing and automated detection* (NBER Working Paper No. 34223). National Bureau of Economic Research. https://doi.org/10.3386/w34223

Jacobs, A. Z., & Wallach, H. (2021). Measurement and fairness. In *Proceedings of the 2021 ACM Conference on Fairness, Accountability, and Transparency* (pp. 375–385). ACM. https://doi.org/10.1145/3442188.3445901

Jeunesse, C. (2025). *Des usages des technologies pour apprendre aux environnements personnels d'apprentissage augmentés par l'intelligence artificielle générative (IAG) : contribution à l'élaboration d'un cadre de recherches relatif aux impacts de l'IAG sur les dispositions des individus à apprendre et l'ingénierie pédagogique* [Habilitation thesis, Université Paris Nanterre]. https://hal.science/tel-05328100

Lardellier, P., & Carré, E. (2025, January 5). L'IA dans l'enseignement supérieur : les leçons de Platon et du mythe de Theuth. *The Conversation*. https://doi.org/10.64628/AAK.5ph5qhp3k

Las Vergnas, O., Jeunesse, C., & Adinda, D. (2025). De la délégation à l'amplification : quand l'IA générative reconfigure nos façons d'apprendre. *Action publique. Recherche et pratiques*, (25), 23–36. https://hal.science/hal-05329510v1

Las Vergnas, O., & Jeunesse, C. (2026, May 27). *Using generative AI as cognitive orthoses: Optimizing LLM uses through controlled dialogues and hybrid agents* [Invited talk]. 3rd Econom'IA Workshop, CY Cergy Paris Université, France. https://economia.sciencesconf.org/resource/page/id/14

Luther, T., Belcher, K., Kimmerle, J., Meurers, D., & Cress, U. (2026). Collaborative knowledge construction with generative AI: Exploring argumentative co-writing processes through n-gram and cluster analysis. *International Journal of Computer-Supported Collaborative Learning, 21*, 619–646. https://doi.org/10.1007/s11412-026-09470-1

Meller, A. (2019, March). Les deux intelligences artificielles : entre prothèse et orthèse. *La Jaune et la Rouge*, (743). https://www.lajauneetlarouge.com/les-deux-intelligences-artificielles-entre-prothese-et-orthese/

Monge Roffarello, A., Calò, T., Scibetta, L., & De Russis, L. (2025). Investigating how computer science researchers design their co-writing experiences with AI. In *Proceedings of the 2025 CHI Conference on Human Factors in Computing Systems* (Article 1215, pp. 1–17). ACM. https://doi.org/10.1145/3706598.3713205

Popper, K. R. (1959). *The logic of scientific discovery*. Hutchinson. (Original work published 1935).

Sarkar, A. (2025). AI could have written this: Birth of a classist slur in knowledge work. In *Extended abstracts of the CHI Conference on Human Factors in Computing Systems* (Article 621, pp. 1–12). ACM. https://doi.org/10.1145/3706599.3716239

Shneiderman, B. (2020). Human-centered artificial intelligence: Reliable, safe & trustworthy. *International Journal of Human–Computer Interaction, 36*(6), 495–504. https://doi.org/10.1080/10447318.2020.1741118

Su, Z., Wang, Y., Wan, H., Zhang, Z., & Luo, M. (2025). HACo-Det: A study towards fine-grained machine-generated text detection under human-AI coauthoring. In *Proceedings of the 63rd Annual Meeting of the Association for Computational Linguistics* (Vol. 1, pp. 22015–22036). ACL. https://doi.org/10.18653/v1/2025.acl-long.1069

Thai, K., Emi, B., Masrour, E., & Iyyer, M. (2026). EditLens: Quantifying the extent of AI editing in text. *International Conference on Learning Representations*. ICLR 2026 proceedings; arXiv:2510.03154 (preprint)

Turnitin. (n.d.). *Using the AI Writing Report*. Turnitin Guides. Retrieved November 28, 2026, from https://guides.turnitin.com/hc/en-us/articles/22774058814093-Using-the-AI-Writing-Report

Weber-Wulff, D., Anohina-Naumeca, A., Bjelobaba, S., Foltýnek, T., Guerrero-Dib, J., Popoola, O., Šigut, P., & Waddington, L. (2023). Testing of detection tools for AI-generated text. *International Journal for Educational Integrity, 19*, Article 26. https://doi.org/10.1007/s40979-023-00146-z

Zeng, Z., Liu, S., Sha, L., Li, Z., Yang, K., Liu, S., Gašević, D., & Chen, G. (2024). Detecting AI-generated sentences in human-AI collaborative hybrid texts: Challenges, strategies, and insights. In *Proceedings of the Thirty-Third International Joint Conference on Artificial Intelligence* (pp. 7545–7553). https://doi.org/10.24963/ijcai.2024/835

Zhang, Q., Gao, C., Chen, D., Huang, Y., Huang, Y., Sun, Z., Zhang, S., Li, W., Fu, Z., Wan, Y., & Sun, L. (2024). LLM-as-a-coauthor: Can mixed human-written and machine-generated text be detected? In *Findings of the Association for Computational Linguistics: NAACL 2024* (pp. 409–436). ACL. https://doi.org/10.18653/v1/2024.findings-naacl.29

Zhao, Z., Fu, J., Wang, W. H., & Zhu, Z. (2026). Towards fine-grained and robust detection of human–AI co-authorship. In *Findings of the Association for Computational Linguistics: EMNLP 2026* (pp. 28981–28998). ACL. https://doi.org/10.18653/2026.findings-emnlp.1365

## References cited in the original article but omitted from its bibliography

European Parliament & Council of the European Union. (2024). Regulation (EU) 2024/1689 laying down harmonised rules on artificial intelligence (Artificial Intelligence Act). *Official Journal of the European Union*. https://eur-lex.europa.eu/eli/reg/2024/1689/oj

Mitchell, E., Lee, Y., Khazatsky, A., Manning, C. D., & Finn, C. (2023). DetectGPT: Zero-shot machine-generated text detection using probability curvature. *Proceedings of Machine Learning Research, 202*, 24950–24962. https://proceedings.mlr.press/v202/mitchell23a.html

Orélien, T. (2026). *C'était ça ou mourir*. Grasset. https://www.grasset.fr/livre/cetait-ca-ou-mourir-9782246847069/

Ouyang, L., Wu, J., Jiang, X., Almeida, D., Wainwright, C. L., Mishkin, P., Zhang, C., Agarwal, S., Slama, K., Ray, A., Schulman, J., Hilton, J., Kelton, F., Miller, L., Simens, M., Askell, A., Welinder, P., Christiano, P. F., Leike, J., & Lowe, R. (2022). Training language models to follow instructions with human feedback. *Advances in Neural Information Processing Systems, 35*, 27730–27744. https://arxiv.org/abs/2203.02155

Swales, J. M. (1990). *Genre analysis: English in academic and research settings*. Cambridge University Press.

# Claiming to Detect AI-Written Texts: An Already Obsolete Conception of Working with LLMs?

By **Olivier Las Vergnas**
Emeritus Professor — Paris Nanterre & Lille University
With the assistance of ChatGPT 6.1-high, in deep-orthesis mode

**Automatic English translation by ChatGPT (GPT-6), November 2026.**
Originally published in French in L'Esprit d'Archimède, ALEA & FST-Lille University, no. 16 (2026). Thanks to them.

**Commercial AI detectors claim to distinguish human-written texts from those generated by large language models (LLMs) by reporting a percentage of supposedly human or non-human sentences. But what does such a count mean when authors increasingly work in dialogue with these systems? Without more nuanced answers, could these detectors impose an already obsolete understanding of intellectual work?**

## Can we detect AI without knowing what we are looking for?

With the spread of generative AI (GenAI), several detector software products have come onto the market. They supposedly address an initially legitimate concern, especially at universities: how can a text be used to assess a person's knowledge or skills when an LLM can produce convincing writing in moments? The difficulty most participants have with this question is plain in the extraordinary controversy surrounding T. Orélien's novel *C'était ça ou mourir* during the literary season (Orélien, 2026). Everyone acted as if the question were whether a work could fall from the sky–or, more precisely, whether there was a threshold of pages produced entirely by an LLM beyond which a literary prize should be withheld.

Yet one of the first questions should be what it means to claim that a text was written "X% by AI." Anyone who uses LLMs can see how diverse and rapidly evolving these systems and their uses are. Between an entirely human work and a text commissioned from a machine lie many forms of work: documentary research, idea generation, discussion of arguments, cross-checking sources, data analysis, automated processing, critical or spelling revision. We have grouped these interactions under prosthesis, orthesis and auxesis (see Figure 1, Box 1 and our earlier articles in this series). LLMs may be involved long before the final sentence, even before a decision to publish. Asking only who polished the final version makes all the other work invisible. In short, using a detector score as the yardstick of human–LLM work reduces this cooperation to naive prosthesis.

## Box 1. Definitions of prosthesis, deep orthesis and unbridled auxesis

Under the logic of prosthesis, the machine substitutes for a function without any intention to exercise control: it writes or analyses in place of the human. Auxesis, by contrast, describes a radical amplification of productive capacities, with a risk of losing control when the tool's excessive output outstrips the human's critical abilities (in the case of unbridled auxesis).

Orthesis, by contrast, supports an activity that remains directed by its user. With an LLM, it can involve sustained questioning, discussion of classifications or critical examination of results. We call *deep orthesis* an explicit, durable organisation of this cooperation, which goes beyond asking for a text or correcting a few sentences. Humans retain the initiative in defining objectives, choosing criteria, exercising oversight and taking responsibility for validation. This approach requires distinguishing LLMs, tools, agents and orchestrators; keeping records of exchanges and consulted documents; breaking tasks into controllable stages; and checking proposals. It also requires resisting the *reflective illusion*: the fluency of responses does not guarantee that a model understands, as a human does, what it asserts or can judge whether it is true.

The concepts of distributed cognition (Hutchins, 1995) and the extended mind (Clark & Chalmers, 1998) help us understand what systematic deep ortheses can offer. Dialogues with an LLM, the documents associated with them, and their revisiting over time can provide spaces in which ideas are confronted. Several working threads may even become "thinking hives": one returns to them, lets them interact indirectly and develops problems over time without conflating them.

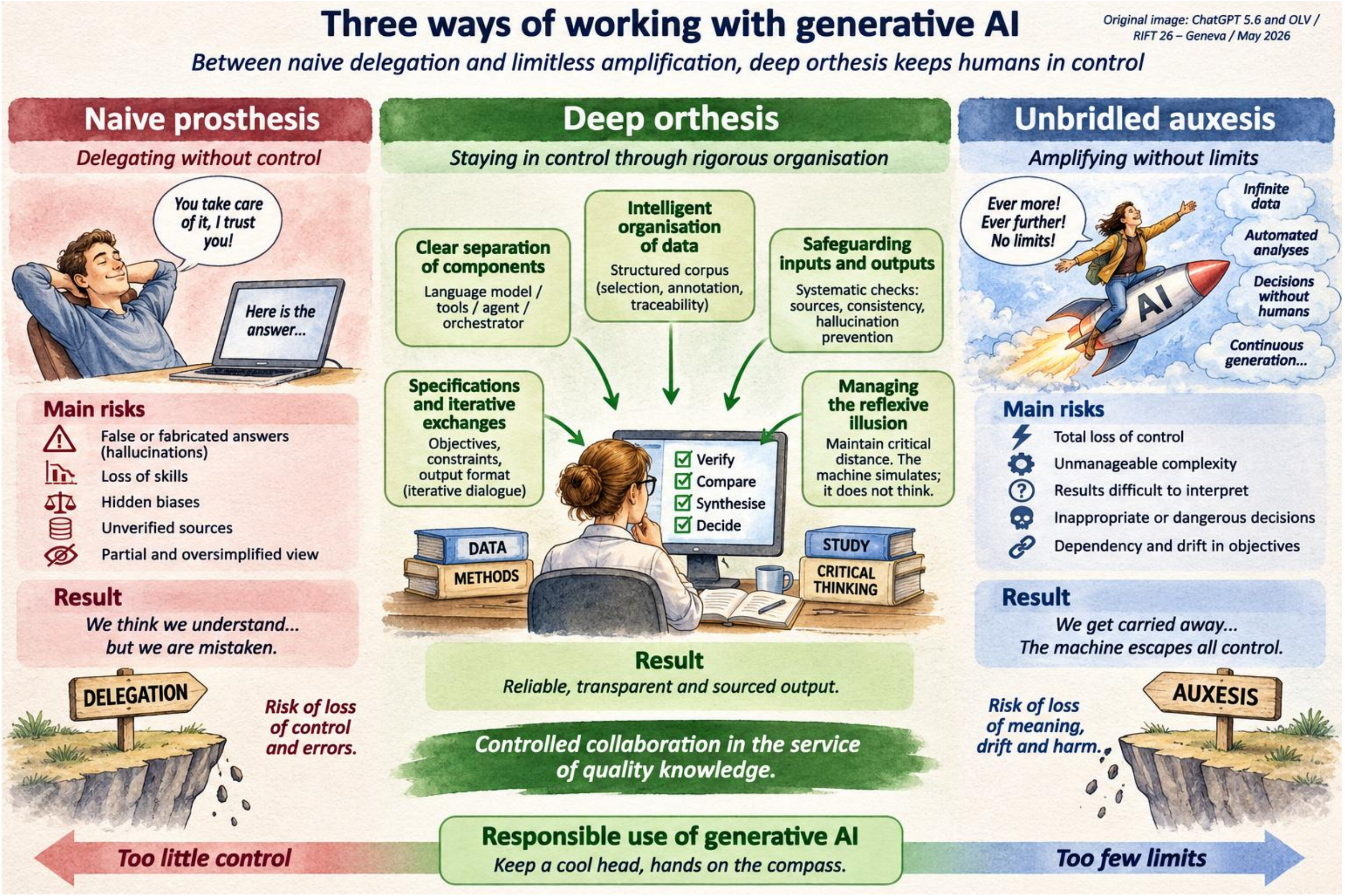

**Figure 1 –** Naive prosthesis, deep orthesis and unbridled auxesis: two extremes and a way to retain human control when working with generative AI (Las Vergnas and ChatGPT; RIFT conference, Geneva, 2026).

## How do commercial AI detectors work?

Commercially available "post hoc" detectors, such as Pangram or Compilatio Magister+, evaluate completed texts. They mainly rely on three operating principles:

**1) Classify à la Pavlov (figuratively speaking).** Discriminative classifiers are trained at scale on texts labelled "human" or "AI." They construct a statistical boundary without reconstructing how the texts were produced. Pangram describes its approach in these terms and also distinguishes "AI-assisted" content. To improve its model, it expands training with difficult-to-classify examples. This is therefore not a conditioned reflex in the strict sense.

**2) Uncover regularities.** Other methods focus on word predictability or stylistic regularities–those familiar linguistic "tics" attributed to LLMs. GLTR, for instance, visualises selected words according to their probability ranks under a language model (Gehrmann et al., 2019). Discriminative classifiers, by contrast, learn to separate two labelled sets of texts as a whole.

**3) Test small variations in the text.** DetectGPT generates variants close to the text under study and estimates their probabilities under a language model. If the original text is more probable than the variants, this is an indication of artificial generation. The point is not to determine whether the original is "better," but to see how small changes affect its probability (Mitchell et al., 2023).

## What happens when post hoc detectors are tested?

While writing this article, we carried out a series of exploratory trials with Pangram. They are neither a representative benchmark nor a general measure of its performance: we wanted to observe how its classifications respond to transformations whose history we knew. The figures below are scores displayed by the tool (see Figure 2), not percentages of words whose provenance has been established.

**Change the order without changing the sentences.** An English text produced by an LLM moves from 89% "AI" to 65% or 79% following simple paragraph permutations. In another case, a GPT-produced French translation of an abstract published in *Science* in 2015 falls from 50% to 32% "AI" when four subheadings are removed without altering the body text. Layout and organisation therefore affect classification.

**Rewrite without changing how the text was commissioned.** Several paraphrases of the same human-written text, produced with GPT, received classifications ranging from "entirely human" to "entirely AI." Conversely, three scientific abstracts generated directly by the model remained classified as 100% "AI" despite changes in syntax, punctuation or sentence length. In our trials, no simple stylistic recipe explains the scores.

**Revise a human-written text.** A 980-word academic passage originally written by a female researcher received an 11% "AI" score. After revision with GPT and additions, a 1,413-word version reached 50%. This comparison involved several changes and cannot isolate their causes. Nevertheless, it reveals one phenomenon: sentences that remained literally identical changed category when their textual context was modified.

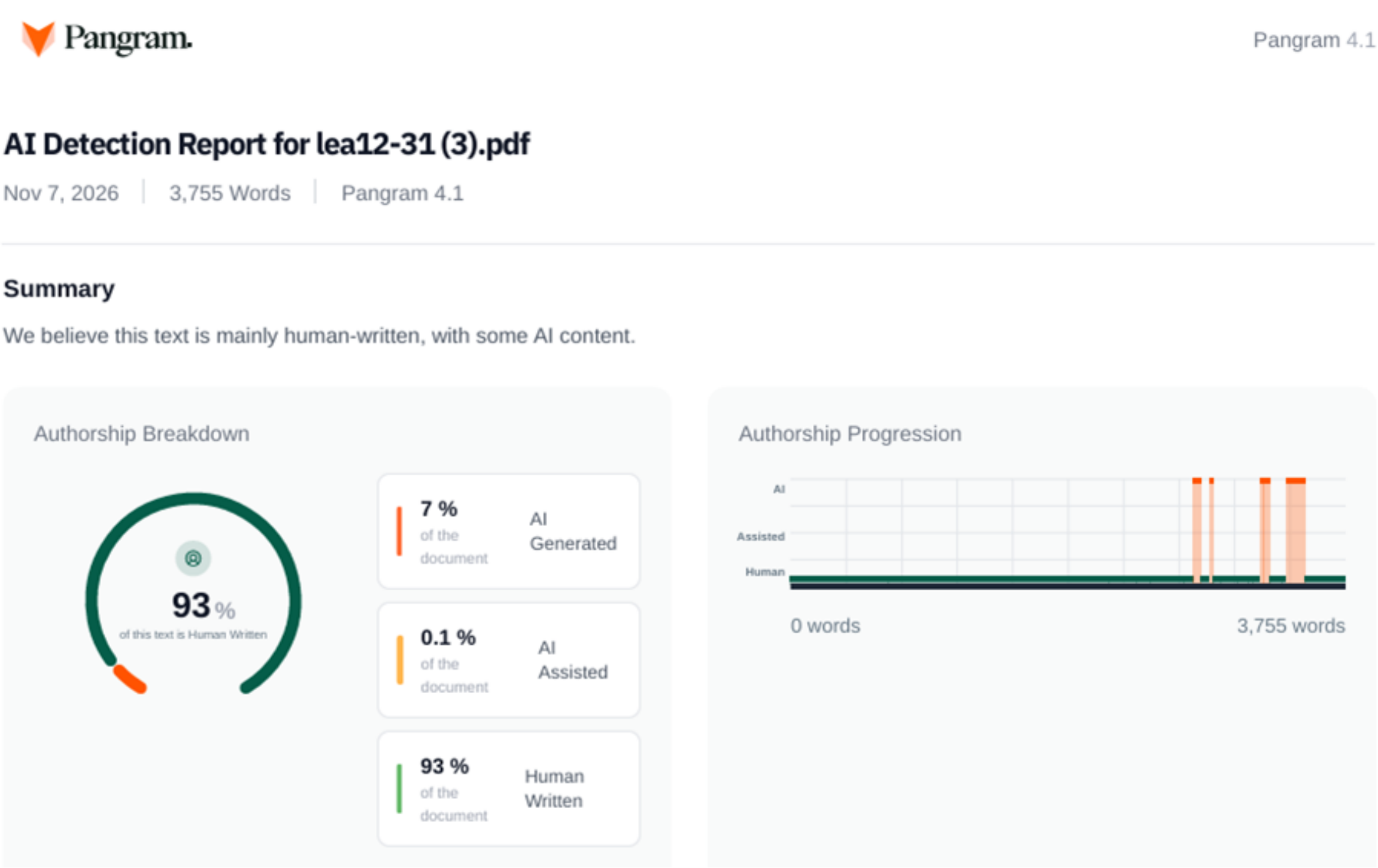

**Figure 2 –** Pangram report on the article published in LEA no. 12 (Las Vergnas, 2024): 93% "human," although many passages are explicitly attributed to ChatGPT-4o in the article.

**Submit a document whose production history is public.** In 2024, *L'Esprit d'Archimède* published an article featuring a text explicitly presented as written by ChatGPT-4o (Las Vergnas, 2024). Pangram classified the complete PDF as 93% "human," although it also contains exchanges, references and editorial material. Passages attributed to ChatGPT even appear in segments classified as human with high confidence (Figure 2). This result does not prove a general failure: it describes the classification of a composite document.

These trials do not measure Pangram's overall error rate. They show that its attributions do not always follow a document's known production history. But what would we learn about reasoning even if every sentence were perfectly attributed?

## Why is it difficult to detect GenAI's contribution after the fact?

Three reasons may seem obvious to humans and machines alike:

**1)** The first has already been mentioned: the co-writing involved in deep orthesis. Seeking to divide a line of reasoning into a "human share" and an "LLM share" ignores the heuristic back-and-forth that characterises this cooperation. A detector might identify machine-written segments; it cannot infer who constructed, controlled and validated the reasoning.

**2)** Alongside the market for detectors, a market for "humanisers" has emerged, promising to make LLM-generated wording less recognisable, especially through paraphrasing. General-purpose rewriting tools, such as QuillBot, may also be used for this purpose.

**3)** Building detection on the difference between human writing and LLM writing assumes that observable indicators of provenance exist. Yet LLMs are trained on human-produced texts! Hence a paradox: through different means, humans and models may both try to bring their texts closer to the same expected norms. Features associated with those norms may then lose some of their discriminative power.

Scientific writing provides a particularly clear case. Much of the work of shaping texts, whether human or artificial, aims to bring them into line with an expected, often highly codified standard specific to disciplines and journals (Hyland, 2004; Swales, 1990). Researchers achieve this through successive revisions; LLMs, trained on human output and adjusted to human preferences, may be asked to meet those same standards (Ouyang et al., 2022). We call the resulting productions "optimised scientific literature" (LSO, from the French *littérature scientifique optimisée*); other genres belong to other families of potentially optimisable texts (FTPO, French acronym). Similar objectives therefore tend to reduce formal differences arising from shared criteria, making systematic detection from the text alone especially difficult.

## How, then, can we identify generative uses of LLMs?

Two approaches move beyond simple post hoc detection. The first is to mark content at its source: Article 50 of the EU AI Act provides for machine-readable marking of generated content, subject to technical conditions and limitations (European Parliament & Council of the European Union, 2024). The other approach is emerging in universities and scientific journals interested in supporting human–LLM co-production in the spirit of deep orthesis.

**1) Mark texts at the source.** Some digital watermarks are inserted during generation, for example by subtly shifting word-selection probabilities; the resulting signature can then be searched for (Kirchenbauer et al., 2023). This technique requires the generator's cooperation and may be weakened by rewriting. But it is only one option among marking and provenance mechanisms: metadata and documentary traces work differently. Above all, marking a text does not establish the intellectual role played by the LLM.

**2) Make deep ortheses visible and traceable.** The second approach is less concerned with policing: instead of prohibiting human–LLM work, it proposes documenting its stages. Dialogues, sources, changes of direction, corrections and validations would form a log of intellectual work. Such a log could be published as open data when rights and confidentiality permit, or made available for auditing. The aim is no longer merely to determine textual origin, but to establish the conditions of human responsibility.

## Conclusion – From detection to tracing cooperation

Nothing presented here invalidates every strategy for recognising the role of LLMs in text production (Figure 3). Our principal aim is to encourage caution regarding the scores and uses of post hoc detectors. Beyond their debatable performance, when these AI detectors are used to judge intellectual work, they help perpetuate an outdated conception of human–machine relations based mainly on substitution or amplification. That conception is inadequate for understanding the practices of reflective cooperation we call "deep orthesis."

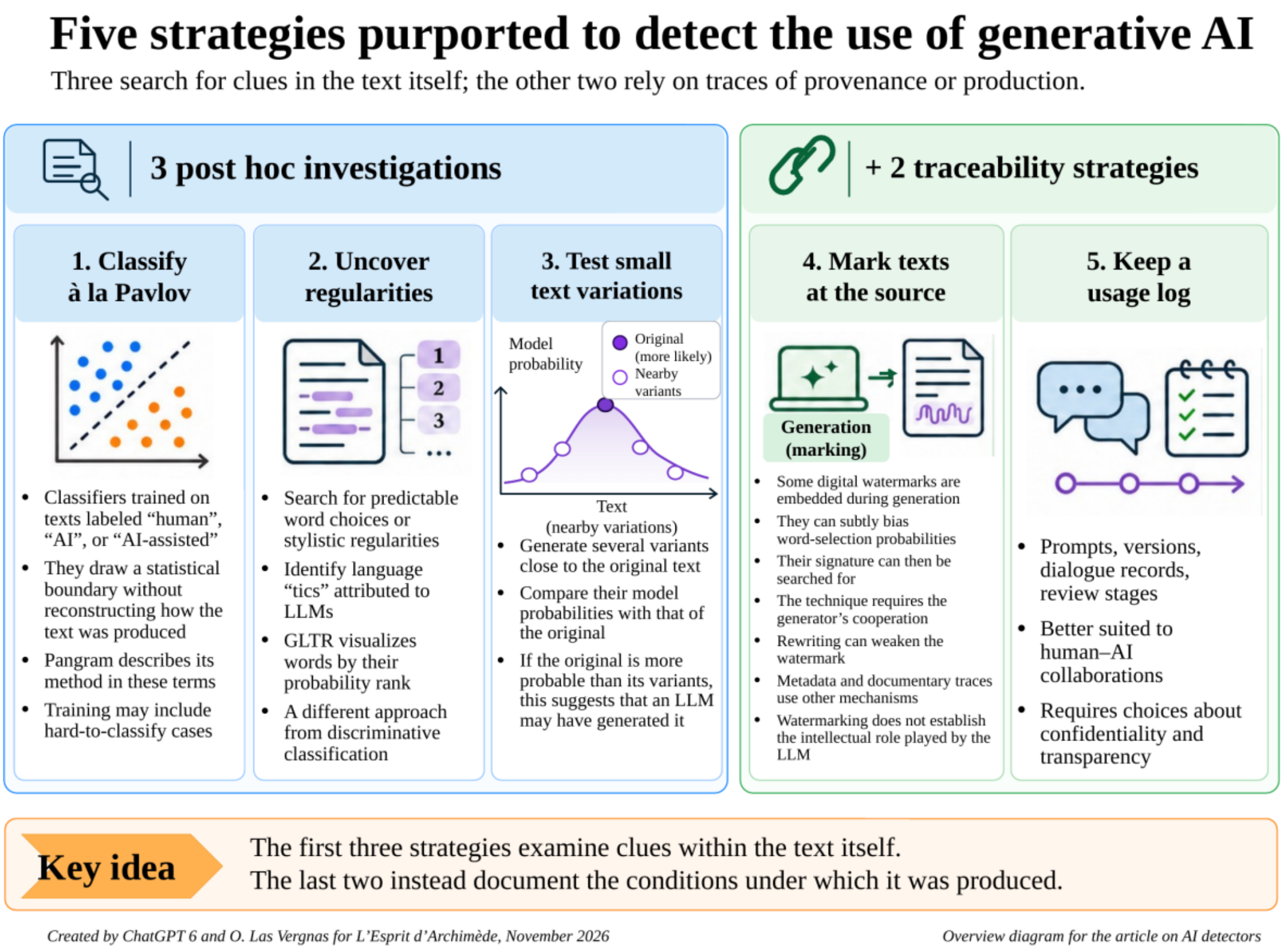

**Figure 3 –** Five strategies purported to detect the use of generative AI: three methods of textual inference and two traceability approaches (O. Las Vergnas and ChatGPT 6, November 2026). Initial prompt, translated from French for traceability: "Could you make an illustration summarising the 3 + 2 detection strategies discussed?"

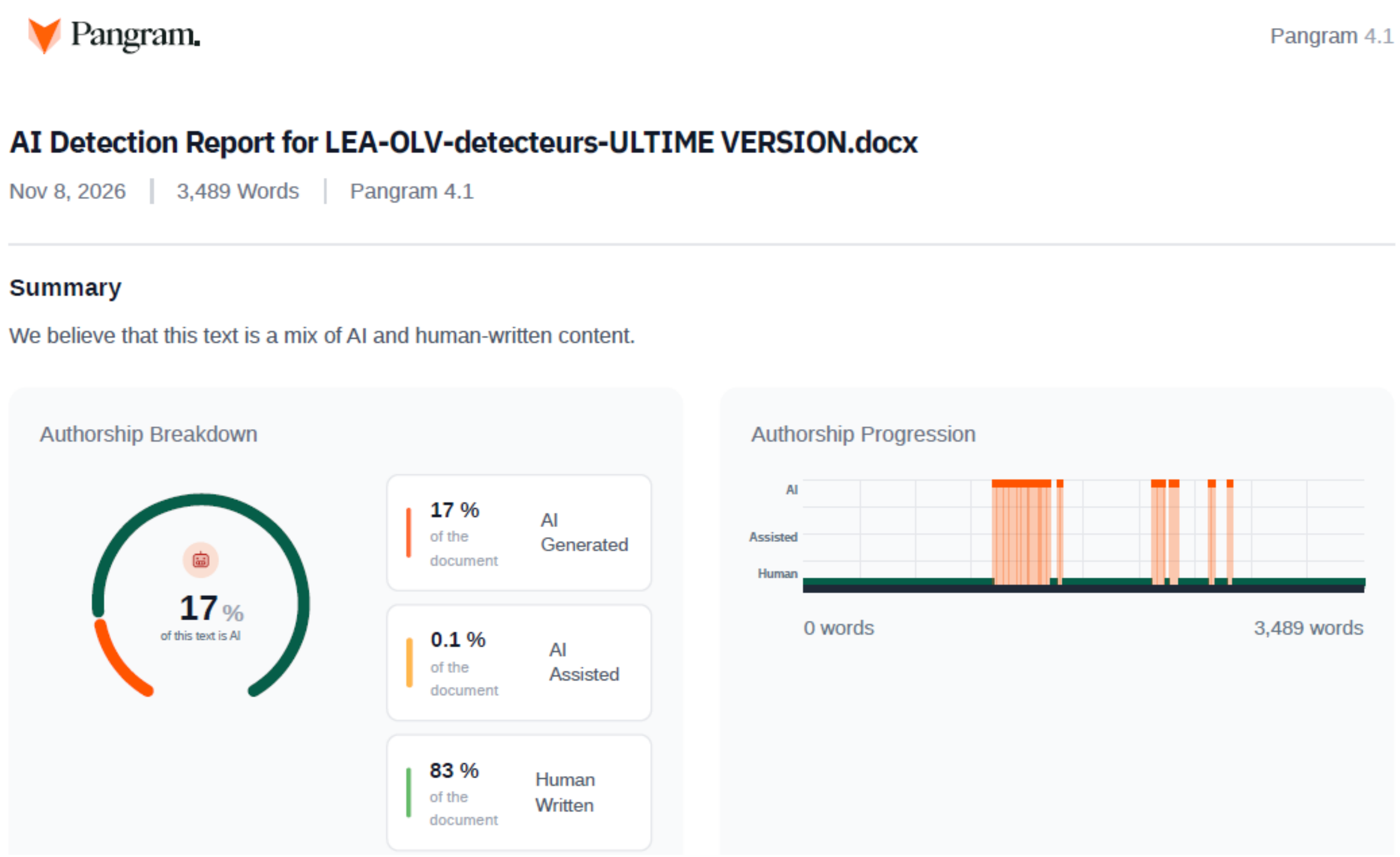

**Figure 4 –** Pangram report on the final version of this article (before adding the figure itself, of course).

## Box 2. How this article was written using deep orthesis

This text grew out of a working dialogue in deep-orthesis mode between Olivier Las Vergnas and GPT-6.1. The researcher defined its central problem: although detectors can be useful in certain contexts, they risk perpetuating a substitution-based view of the human–AI relationship. He proposed the concepts to present, carried out the Pangram trials, supplied their reports and discussed the interpretations. The model contributed technical perspectives, possible outlines, formulations and revisions. Several passages were revised, challenged or clarified over successive exchanges. We maintained the distinction between observed results and the hypotheses they suggest. Ultimately, the human author rewrote the final version to simplify it, reorganise it and focus it on the essential argument. He then asked the LLM to correct the syntax and posed the following question: “Could you give me your assessment of the substance and coherence of this version and, if appropriate, the three improvements you would make?” He incorporated four of the improvements suggested by the LLM. The Pangram score consequently rose from 10% “AI” to 23% “AI,” then fell to 17% after the final typographical corrections (Figure 4). Such co-writing does not make the LLM a responsible author: the choice of arguments, verification of sources and editorial decisions remain under human control. The experience illustrates our argument: the text’s production history cannot be inferred from its final form. The main discussion thread underpinning this deep orthesis is available as open data on Nakala at https://nakala.fr/10.34847/nkl.afa74m7q (version-specific reference in the original French edition).

## References

Clark, A., & Chalmers, D. (1998). The extended mind. *Analysis, 58*(1), 7–19. https://doi.org/10.1093/analys/58.1.7

Gehrmann, S., Strobelt, H., & Rush, A. M. (2019). GLTR: Statistical detection and visualization of generated text. In *Proceedings of the 57th Annual Meeting of the Association for Computational Linguistics: System Demonstrations* (pp. 111–116). Association for Computational Linguistics. https://doi.org/10.18653/v1/P19-3019

Hutchins, E. (1995). *Cognition in the wild.* MIT Press. https://doi.org/10.7551/mitpress/1881.001.0001

Hyland, K. (2004). *Disciplinary discourses: Social interactions in academic writing* (Michigan Classics ed.). University of Michigan Press. https://doi.org/10.3998/mpub.6719

Kirchenbauer, J., Geiping, J., Wen, Y., Katz, J., Miers, I., & Goldstein, T. (2023). A watermark for large language models. *Proceedings of Machine Learning Research, 202*, 17061–17084. https://proceedings.mlr.press/v202/kirchenbauer23a.html

Las Vergnas, O. (2024). Rédaction automatique d’article par une intelligence artificielle : Points de vigilance. Un article rédigé par ChatGPT4o sur la suggestion d’Olivier Las Vergnas [Automated article writing by artificial intelligence: Points of caution. An article written by ChatGPT-4o at the suggestion of Olivier Las Vergnas]. *L’Esprit d’Archimède*, (12), 31–35. https://alea.univ-lille.fr/revue.php?article=31&idn=12

**Full version-specific Nakala link:**

https://nakala.fr/10.34847/nkl.afa74m7q#4f0a00bac6216e90f677b332026534b1db42f691

*Automatic English translation by ChatGPT (GPT-6), 17 November 2026.*

# Prétendre détecter des textes écrits par une IA : une conception déjà obsolète du travail avec les LLM ?

Par **Olivier Las Vergnas**

Professeur émérite des universités Avec l'aide de ChatGPT 6.1-high, en mode orthèse profonde

**Les détecteurs d'IA disponibles sur le marché prétendent distinguer les textes humains de ceux qu'engendrent les modèles de langage (LLM) en donnant un pourcentage de phrases humaines ou non. Mais que signifie une telle façon de compter alors que les auteurs travaillent maintenant de plus en plus en dialogue avec ces systèmes ? En l'absence de réponses plus fines, les usages de ces détecteurs ne risquent-ils pas d'imposer une vision déjà obsolète du travail intellectuel ?**

## Peut-on détecter de l'IA sans savoir ce que l'on cherche ?

Avec la diffusion des IA génératives (IAGen), plusieurs logiciels détecteurs ont été mis sur le marché. Ils sont censés répondre à une préoccupation a priori légitime, notamment à l'université : comment apprécier à travers un texte les connaissances ou compétences d'un humain lorsqu'un LLM peut produire en quelques instants un texte convaincant ? La non-maîtrise de cette question par la plupart des protagonistes est facile à voir lors d'épisodes rocambolesques comme ceux du feuilleton de la rentrée littéraire créé par le roman « C'était ça ou mourir » de T. Orélien (Orélien, 2026). Tout le monde a fait comme s'il s'agissait de savoir si oui ou non une œuvre pouvait tomber du ciel ou, plus exactement, s'il existait un seuil de pages entièrement pondues par un LLM au-delà duquel refuser un prix littéraire.

Or, une des premières questions devrait être de se demander ce que peut bien signifier l'affirmation selon laquelle un texte serait écrit à X % par une IA ? En fait, dès lors que l'on est soi-même utilisateur de LLM, on voit bien à quel point les systèmes et usages sont diversifiés et surtout en pleine évolution. Entre une œuvre entièrement humaine et un texte commandé à une machine, se déploient des formes multiples de travail : recherche documentaire, ruche à idées, discussion des arguments, confrontation de sources, analyse de données, automatisation de traitement, révision critique ou orthographique : autant de façons d'interagir que nous avons déjà regroupées en prothèse, orthèse ou auxèse (voir la figure 1, l'encadré 1 et nos précédents articles dans cette rubrique). Les LLM peuvent être mobilisés longtemps avant la phrase finale, voire avant même le projet de publier... Se limiter à chercher à savoir qui a ciselé la dernière version rend tout le reste du travail invisible. En bref, faire du score d'un détecteur l'étalon du travail humain–LLM, c'est réduire cette coopération à la seule prothèse naïve.

## Encadré 1 :Rappel des définitions de prothèse, orthèse profonde et auxèse débridée

Dans la logique de prothèse, la machine se substitue à une fonction, sans volonté de contrôle : elle rédige ou analyse à la place de l'humain. L'auxèse décrit, elle, une amplification radicale des capacités de production avec le risque de perte de contrôle par dépassement des compétences critiques de l'humain, submergé par une productivité démesurée de l'outil (dans le cas de l'auxèse débridée).

L'orthèse soutient, au contraire, une activité dont l'utilisateur garde le pilotage. Avec un LLM, elle peut prendre la forme d'un questionnement suivi, d'une discussion sur les classifications ou d'un examen critique de résultats. Nous nommons orthèse profonde une organisation explicite et durable de cette coopération, qui ne se limite pas à demander un texte ou à corriger quelques phrases. L'humain conserve l'initiative des finalités, le choix des critères, les compétences de contrôle et la responsabilité des validations. Cette démarche demande de distinguer LLM, outils, agents et orchestrateurs ; de garder des traces des échanges et des documents mobilisés ; de découper les tâches en étapes contrôlables ; et de soumettre les propositions à des vérifications. Elle suppose aussi de résister à l'illusion réflexive : la fluidité des réponses ne garantit pas que le modèle comprenne, à la manière d'un humain, ce qu'il affirme ou qu'il puisse en juger la vérité.

Les notions de cognition distribuée (Hutchins, 1995) et d'esprit étendu (Clark & Chalmers, 1998) aident à mieux comprendre ce que permet la systématisation des orthèses profondes. Les dialogues avec un LLM, leurs documents associés et leur reprise dans le temps peuvent constituer des lieux de confrontation d'idées. Plusieurs fils de travail deviennent même des « ruches à penser » : on y revient, on les fait dialoguer indirectement, on y mûrit des problèmes sans les confondre.

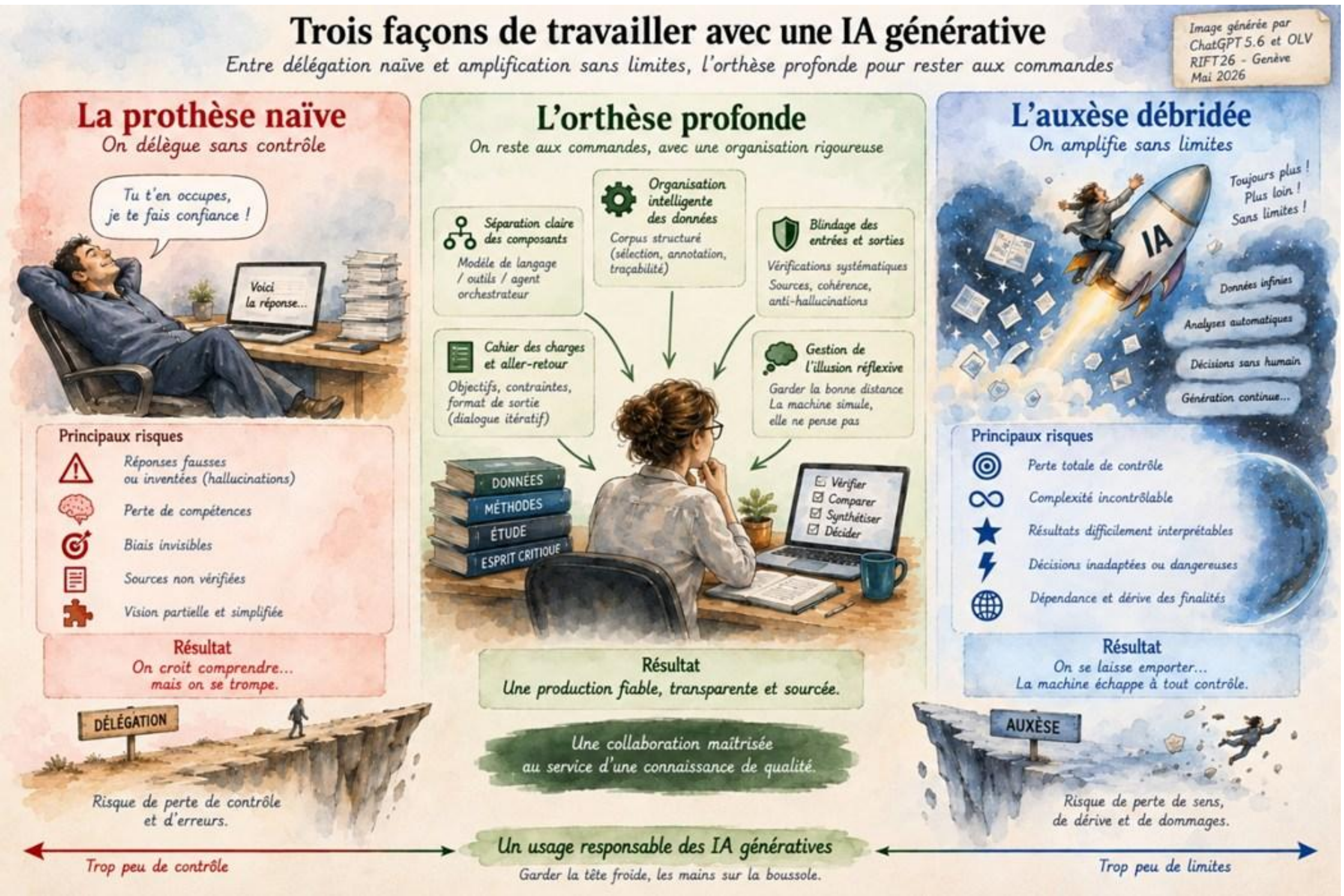

Figure 1 — Prothèse naïve, orthèse profonde et auxèse débridée : trois façons de travailler avec une IA générative (Las Vergnas et ChatGPT, colloque RIFT, Genève, 2026).

## Comment marchent les détecteurs du commerce ?

Sur le marché, on trouve des détecteurs « a posteriori » travaillant à évaluer des textes terminés, comme Pangram ou Compilatio Magister+. Ils déploient principalement trois principes de fonctionnement :

1) Discriminer à la Pavlov (façon de parler). Les classificateurs discriminants sont entraînés massivement sur des textes étiquetés « humain » ou « IA » : ils construisent une frontière statistique, sans reconstituer la genèse des textes. Pangram revendique cette approche et distingue aussi « assisté par IA ». Pour améliorer son modèle, il enrichit notamment l'entraînement avec des exemples difficiles à classer. Rien d'un réflexe conditionné au sens strict, donc.

2) Démasquer des régularités. D'autres méthodes s'intéressent à la prévisibilité des mots ou à certaines régularités stylistiques, ces fameux « tics de langage » attribués aux LLM. GLTR visualise, par exemple, les mots choisis selon leur rang de probabilité pour un modèle de langage (Gehrmann et al., 2019). À la différence de cette démarche, les classificateurs discriminants apprennent globalement à départager deux ensembles de textes étiquetés.

3) Tester de petites variations du texte. DetectGPT produit des variantes proches du texte étudié et estime leurs probabilités respectives d'avoir été finalement celle qui aurait été proposée par le LLM. Si le texte initial a plus de probabilité que toutes les variantes, c'est un indice de génération artificielle. Il ne s'agit pas de vérifier si l'original est «meilleur », mais de voir comment de petites retouches font varier sa probabilité (Mitchell et al., 2023).

## Que constate-t-on lorsqu'on teste les détecteurs a posteriori ?

À l'occasion de l'écriture de cet article, nous avons conduit une série d'essais exploratoires avec Pangram. Il ne s'agit ni d'un banc d'essai représentatif ni d'une mesure générale de ses performances : nous voulions observer comment ses classements réagissent à des transformations dont nous connaissions l'histoire. Les chiffres qui suivent sont les scores affichés par l'outil (voir notamment la figure 2), et non des pourcentages de mots dont l'origine serait établie.

Changer l'ordre, sans toucher aux phrases. Un texte anglais produit par un LLM passe de 89 % « IA » à 65 % ou 79 % après de simples permutations de paragraphes. Ailleurs, la traduction française par GPT d'un résumé publié dans Science en 2015 descend de 50 % à 32 % « IA » lorsqu'on retire quatre intertitres, sans modifier le corps du texte. L'organisation matérielle intervient donc dans la classification.

Réécrire, sans changer l'origine de la commande. Plusieurs paraphrases d'un même texte humain, réalisées par GPT, ont reçu des attributions allant de « totalement humain » à « totalement IA ». À l'inverse, trois résumés scientifiques générés directement par le modèle sont restés classés à 100 % « IA » malgré des changements de syntaxe, de ponctuation ou de longueur des phrases. Il n'existe pas, dans nos essais, de recette stylistique simple expliquant les scores.

Réviser un texte humain. Un passage universitaire de 980 mots, rédigé initialement par une chercheuse, obtient 11 % « IA ». Après révision avec GPT et ajouts, il atteint 50 % pour 1 413 mots. Cette comparaison cumule plusieurs modifications ; elle ne permet pas d'en isoler les causes. Elle révèle néanmoins un phénomène : des phrases restées littéralement identiques changent de catégorie lorsque leur contexte rédactionnel est modifié.

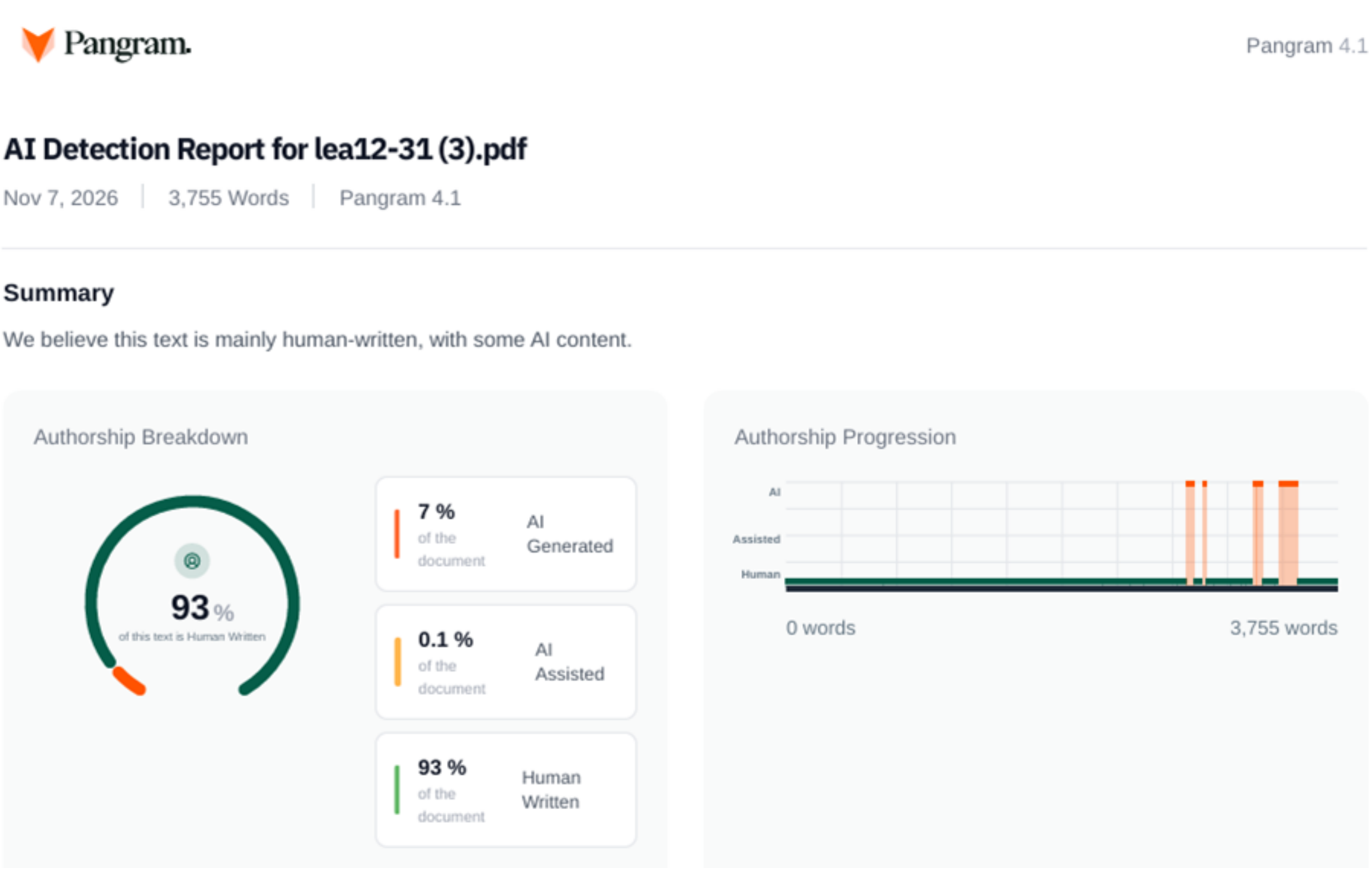

Figure 2 — Rapport Pangram sur l'article paru dans LEA n° 12 (Las Vergnas, 2024) : 93 % « humain » alors que de nombreux passages sont attribués explicitement à ChatGPT-4o dans l'article.

Soumettre un document dont la genèse est publiée. En 2024, *L'esprit d'Archimède* publiait un article mettant en scène un texte annoncé comme rédigé par ChatGPT-4o (Las Vergnas, 2024). Pangram classe à 93 % « humain » le PDF complet, qui comprend aussi échanges, références et éléments éditoriaux. Des passages attribués à ChatGPT figurent même dans des segments classés humains avec confiance élevée (figure 2). Ce résultat ne prouve pas une défaillance générale : il décrit le classement d'un document composite.

Ces essais ne mesurent pas le taux d'erreur général de Pangram. Ils montrent que ses attributions ne suivent pas toujours la genèse connue d'un document. Mais qu'apprendrions-nous sur le raisonnement si chaque phrase était parfaitement attribuée ?

## Pourquoi est-il difficile de détecter a posteriori la part de l'IAGen ?

Trois raisons peuvent paraître évidentes, aussi bien pour les humains que pour une machine :

1) Nous avons déjà évoqué la première, qui concerne les coécritures en orthèse profonde. Vouloir répartir un raisonnement entre « part humaine » et « part LLM » méconnaît les allers-retours heuristiques qui caractérisent cette coopération. Un détecteur peut éventuellement identifier des segments de rédaction artificielle ; il ne peut en déduire qui a construit, contrôlé et validé le raisonnement.

2) Parallèlement au marché des détecteurs s'est développé celui des « humaniseurs », qui promettent de rendre moins reconnaissables les formulations produites par LLM, notamment au moyen de paraphrases. Des outils généralistes de réécriture, comme QuillBot, peuvent aussi être employés dans ce but.

3) Fonder une détection sur la différence entre écriture humaine et écriture de LLM suppose qu'il existe des indices observables de leur provenance. Or les LLM sont précisément entraînés sur des productions humaines ! D'où un paradoxe : humains et modèles peuvent chercher, par des voies différentes, à rapprocher leurs textes des mêmes normes attendues. Les caractéristiques associées à ces normes risquent alors de perdre une partie de leur pouvoir discriminant.

L'écriture scientifique en offre un cas privilégié. L'essentiel du travail de mise en forme, tant humain qu'artificiel, consiste à rapprocher les textes d'une norme attendue, souvent très codifiée, propre aux disciplines et aux revues (Hyland, 2004 ; Swales, 1990). Les chercheurs y parviennent par des révisions successives ; les LLM, entraînés sur des productions humaines et ajustés selon des préférences humaines, peuvent être sollicités pour viser ces mêmes normes (Ouyang et al., 2022). Nous appelons « littérature scientifique optimisée » (LSO) les productions ainsi travaillées ; d'autres genres relèvent d'autres familles de textes potentiellement optimisables (FTPO). La similitude des objectifs tend ainsi à réduire les différences formelles liées à ces critères communs, rendant toute perspective de détection systématique, à partir du seul texte, particulièrement délicate.

## Mais alors comment repérer les usages génératifs des LLM ?

Deux pistes sortent de la simple détection a posteriori. La première consiste à marquer les contenus à la source : l'article 50 du règlement européen sur l'IA prévoit un marquage lisible par machine des contenus générés, selon certaines conditions et limites techniques (Parlement européen & Conseil de l'Union européenne, 2024). L'autre commence à se mettre en place dans des universités et des revues scientifiques trouvant intéressant d'accompagner le déploiement des coproductions humain-LLM dans la logique de l'orthèse profonde.

1) Marquer les textes à la source. Certains filigranes numériques (watermarks) sont insérés lors de la génération, par exemple en infléchissant discrètement les probabilités de choix des mots ; on peut ensuite en rechercher la signature (Kirchenbauer et al., 2023). Cette technique suppose la coopération du générateur et peut être fragilisée par une réécriture. Elle n'est toutefois qu'une possibilité parmi les dispositifs de marquage ou de provenance : métadonnées et traces documentaires relèvent d'autres mécanismes. Surtout, marquer un texte n'établit pas le rôle intellectuel joué par le LLM.

2) Rendre visibles et traçables les orthèses profondes. La seconde piste est moins policière : plutôt que de prohiber le travail humain–LLM, elle propose d'en documenter les étapes. Dialogues, sources, bifurcations, corrections et validations constitueraient un journal du travail intellectuel. Celui-ci pourrait être publié en open data lorsque les droits et la confidentialité le permettent, ou rester accessible dans un cadre d'audit. On ne cherche plus seulement une origine textuelle, mais les conditions d'une responsabilité humaine.

## Conclusion — De la détection à la traçabilité des coopérations

Ce que nous exposons ici n'invalide pas toutes les stratégies de reconnaissance de la place des LLM dans la production textuelle (fig. 3). Nous voulons surtout inviter à la prudence devant les scores et les rôles des détecteurs a posteriori. Au-delà de leurs performances discutables, ces détecteurs d'IA, lorsqu'ils servent à juger le travail intellectuel, contribuent à entretenir une représentation rétrograde des rapports humain–machine, essentiellement fondée sur la substitution ou l'amplification. Cette représentation devient inadéquate pour comprendre les pratiques de coopération réflexive que nous appelons « orthèse profonde ».

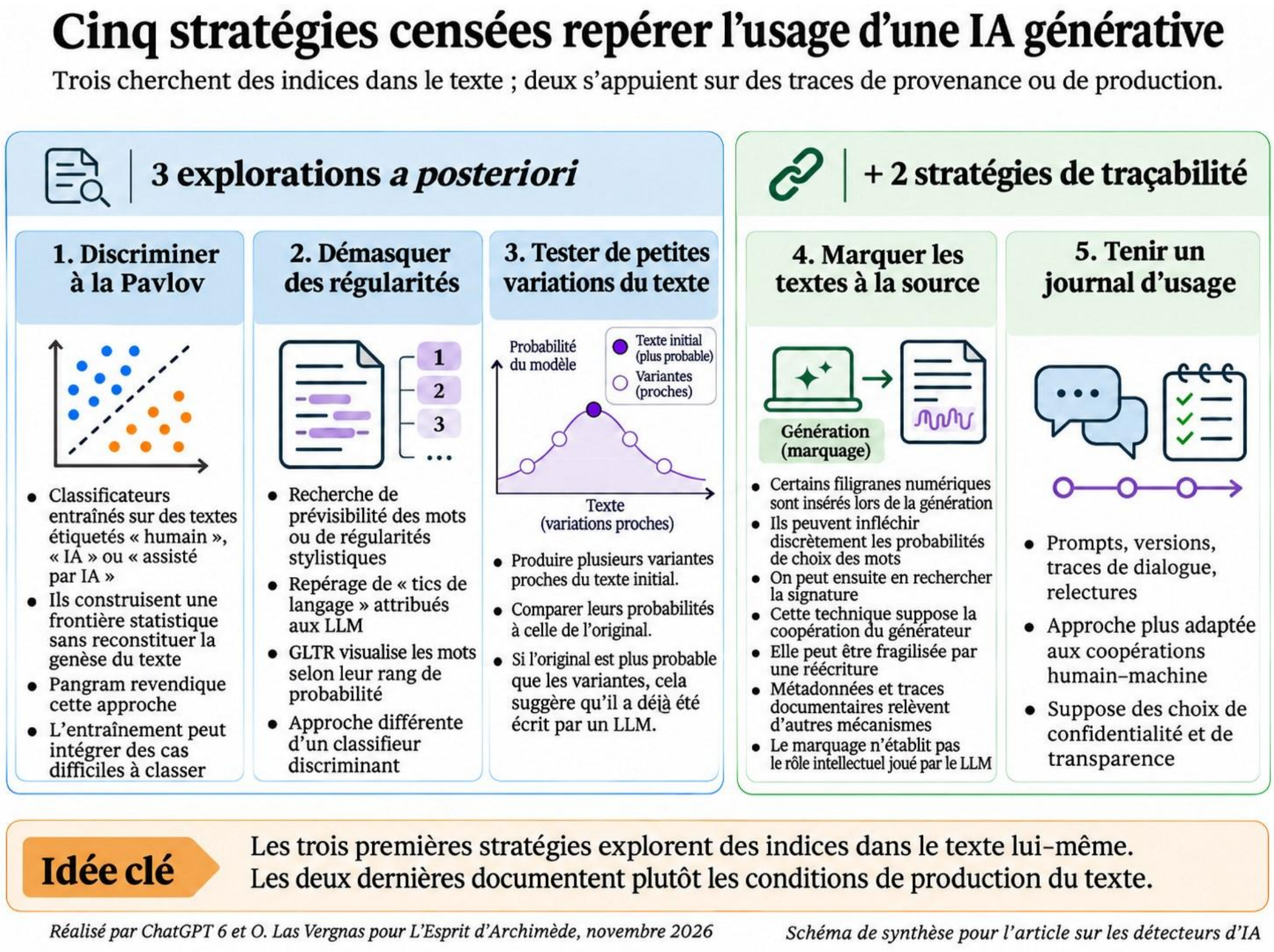

Figure 3— Cinq stratégies pour repérer l'usage d'une IA générative : trois méthodes d'inférence textuelle et deux approches de traçabilité (O. Las Vergnas et ChatGPT 6, novembre 2026). Prompt initial reproduit pour la traçabilité : « *est ce que tu peux me faire une illustration qui résumerait les 3 + 2 stratégies de détection évoquées ?* »

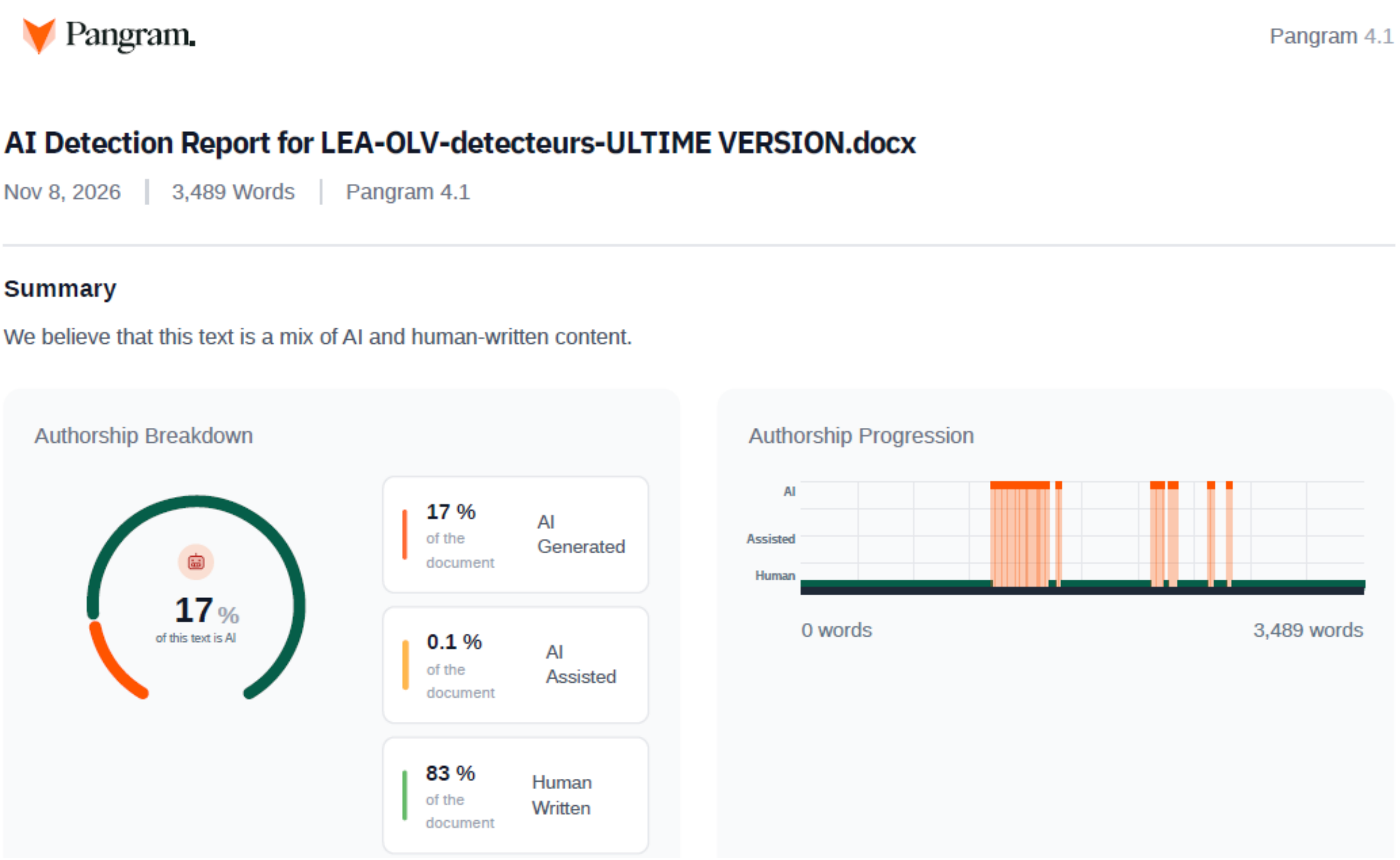

Figure 4 — Rapport Pangram sur la version ultime du présent article (avant intégration de cette figure, bien sûr).

## Encadré 2 : Comment cet article a été écrit en orthèse profonde

Ce texte est issu d'un dialogue de travail en orthèse profonde entre Olivier Las Vergnas et GPT-6.1. Le chercheur en a fixé la problématique centrale : les détecteurs, utiles dans certains contextes, risquent de reconduire une vision substitutive de la relation humain–IA. Il a proposé les notions à présenter, conduit les essais avec Pangram, soumis les rapports et discuté leurs interprétations. Le modèle a contribué aux mises en perspective techniques, aux propositions de plan, aux formulations et aux révisions. Plusieurs passages ont été repris, contestés ou précisés au fil des échanges. Nous avons conservé la distinction entre les résultats observés et les hypothèses qu'ils suggèrent. Finalement, l'humain a repris l'écriture finale pour la simplifier, la réorganiser et la recentrer sur l'argumentaire essentiel. Il a ensuite demandé au LLM de corriger la syntaxe, puis lui a posé la question suivante : « Peux-tu me donner ton avis sur le fond, la cohérence de cette version et, le cas échéant, les trois améliorations que tu voudrais effectuer ? » Il a intégré quatre améliorations proposées par le LLM. Le score Pangram est alors passé de 10 % « IA » à 23 % « IA » puis retombé finalement à 17% après les dernières corrections typographiques (voir figure 4). Cette co-rédaction ne fait pas du LLM un auteur responsable : la sélection des arguments, la vérification des sources et les décisions éditoriales demeurent sous contrôle humain. L'expérience illustre notre propos : sa genèse ne se laisse pas déduire du texte final. Le fil principal de discussion ayant permis cette orthèse profonde est disponible, selon les principes de l'open data sur Nakala à l'adresse https://nakala.fr/10.34847/nkl.afa74m7q#4f0a00bac6216e90f677b332026534b1db42f691

## Références bibliographiques

Clark, A., & Chalmers, D. (1998). The extended mind. *Analysis, 58*(1), 7–19. https://doi.org/10.1093/analys/58.1.7

Gehrmann, S., Strobelt, H., & Rush, A. M. (2019). GLTR: Statistical detection and visualization of generated text. In *Proceedings of the 57th Annual Meeting of the Association for Computational Linguistics: System Demonstrations* (pp. 111–116). Association for Computational Linguistics. https://doi.org/10.18653/v1/P19-3019

Hutchins, E. (1995). *Cognition in the wild*. MIT Press. https://doi.org/10.7551/mitpress/1881.001.0001

Hyland, K. (2004). *Disciplinary discourses: Social interactions in academic writing* (Michigan Classics ed.). University of Michigan Press. https://doi.org/10.3998/mpub.6719

Kirchenbauer, J., Geiping, J., Wen, Y., Katz, J., Miers, I., & Goldstein, T. (2023). A watermark for large language models. *Proceedings of Machine Learning Research, 202*, 17061–17084. https://proceedings.mlr.press/v202/kirchenbauer23a.html

Las Vergnas, O. (2024). Rédaction automatique d'article par une intelligence artificielle : Points de vigilance. Un article rédigé par ChatGPT4o sur la suggestion d'Olivier Las Vergnas. *L'Esprit d'Archimède*, (12), 31–35. https://alea.univ-lille.fr/revue.php?article=31&idn=12